\documentclass[11pt]{article}
\usepackage{moriond,epsfig}

\def\Journal#1#2#3#4{{#1} {\bf #2}, #3 (#4)}

\def\be{\begin{equation}}
\def\ee{\end{equation}}
\def\bea{\begin{eqnarray}}
\def\eea{\end{eqnarray}}

\begin{document}
\vspace*{4cm}
\title{$\eta$ PHYSICS AND $\phi$ RADIATIVE DECAYS at KLOE}

\author{THE KLOE COLLABORATION: \scriptsize F.AMBROSINO, A.ANTONELLI, M.ANTONELLI, C.BACCI, P.BELTRAME, G.BENCIVENNI, S.BERTOLUCCI, C.BINI, C.BLOISE, S.BOCCHETTA, V.BOCCI, F.BOSSI, P.BRANCHINI, R.CALOI, P.CAMPANA, G.CAPON, T.CAPUSSELA, F.CERADINI, S.CHI, G.CHIEFARI, P.CIAMBRONE, E.DE LUCIA, A.DE SANTIS, P.DE SIMONE, G.DE ZORZI, A.DENIG, A.DI DOMENICO, C.DI DONATO, S.DI FALCO, B. DI MICCO, A.DORIA, M.DREUCCI, G.FELICI, A.FERRARI, M.L.FERRER, G.FINOCCHIARO, S.FIORE, C.FORTI, P.FRANZINI, C.GATI, P.GAUZZI, S.GIOCANNELLA, E.GORINI, E.GRAZIANI, M.INCAGLI, W.KLUGE, V.KULIKOV, F.LACAVA, G.LANFRANCHI, J.LEE-FRANZINI, D.LEONE, M.MARTINI, P.MASSARTTI, W.MEI, S.MEOLA, S.MISCETTI, M.MOULSON, S.M\"ULLER, F.MURTAS, M.NAPOLITANO, F.NGUYEN, M.PALUTAN, E.PASQUALUCCI, A.PASSERI, V.PATERA, F.PERFETTO, M.PRIMAVERA, P.SANTANGELO, G.SARACINO, B.SCIASCIA, A.SCIUBBA, F.SCURI, I.SFILIGOI, T.SPADARO, M.TESTA, L.TORTORA, P.VALENTE, B.VALERIANI, G.VENANZONI, R.VERSACI, G.XU.}

\author{presented by: B. DI MICCO}

\address{Universit\`a degli Studi di Roma Tre, I.N.F.N. Roma III}

\maketitle\abstracts{
Here we present KLOE results on  the $\phi$ meson decays  in $\pi^0$ $\pi^0$ $\gamma$, $\pi^+$ $\pi^-$ $\gamma$ and $\eta$ $\pi^0$ $\gamma$, 
the measurement of the ratio $Br(\phi \to \eta' \gamma)/Br(\phi \to \eta \gamma)$ with the estimate of the $\eta'$ gluonium content and the measurement of the $\eta$ mass.} 
\section{Introduction}
The KLOE experiment\cite{KLOE} is performed at the Frascati $\phi$ factory DA$\Phi$NE\cite{dafne}.
DA$\Phi$NE is a high luminosity $e^+$,$e^-$ collider working at $\sqrt{s} \sim 1020$ MeV, corresponding to the  $\phi$ meson mass. 
In the whole period of data taking ($2001 - 2006$) KLOE has collected an integrated luminosity of 2.5 fb$^{-1}$, corresponding to about 8 billions of $\phi$ produced and
$~ 100$ millions of $\eta$ mesons through the electromagnetic decay $\phi \to \eta \gamma$. The main part of these events are stored on tape,  the trigger efficiency ranging from 95\% to 100\%. The analyses described here are performed on the data collected in the years 2001-2002 corresponding to about 1/5 of all KLOE statistics.

\section{$\phi$ decays to scalars.}
\subsection{$\phi \to f_0 \gamma \to \pi^0 \pi^0 \gamma$, $\phi \to f_0 \gamma \to \pi^+ \pi^- \gamma$}
KLOE has recently published\cite{f0p0p0g} a study of the Dalitz plot of the decay $\phi \to \pi^0 \pi^0 \gamma$.
In order to fit the Dalitz plot we have used two phenomenological models, one based on a Kaon-loop approach \cite{Achasov}, KL in the following, and one
based on a point-like coupling between the $\phi$ and the scalars ($f_0$,$a_0$), ``no structure'' (NS)\cite{Maiani} in the following . In order to get an
acceptable $\chi^2$ value for the fit, the $\sigma$ is needed in the KL approach whose parameters we have fixed at $M_\sigma = 462$ MeV/c$^2$ and 
$\Gamma_\sigma = 300$ MeV/c$^2$, the P$(\chi^2)$ varies from $10^{-4}$ without $\sigma$ to 14\% with $\sigma$. Extracting the scalar part of the amplitude from the fitted model we compute $Br(\phi \to f_0 \gamma \to \pi^0 \pi^0 \gamma) = \left [1.07^{+0.01}_{-0.04}(fit)^{+0.04}_{-0.02}(syst)^{+0.06}_{-0.05}(mod) \right] \times 10^{-4}$.

Isospin symmetry relates the $\pi^0 \pi^0$ state to the $\pi^+$ $\pi^-$ state. 
The background channels in the two final states are instead very different. In both cases there is large interference between the signal
and the background, so it is very important to study both final states in order to
cross check each other. The study of the $f_0$ in the $\pi^+ \pi^-$ final state has been published in the paper by KLOE\cite{f0pppm}, 
in this analysis an earlier version of the KL \cite{Achasov_bef} and the NS model has been used.
Both mass values $m_{f_0}$, and  couplings $R = g^2_{fKK}/g^2_{f\pi^+ \pi^-}$ are in good agreement when using the KL fit while they show significant
deviation in the NS fit. In general a large fit instability is seen using the NS model. The value of the Br, 
evaluated as integral of the scalar amplitude, is $\sim 2.1 - 2.4 \times 10^{-4}$, in agreement with the isospin expectation 
$Br(f_0 \to \pi^+ \pi^-) \sim 2 Br(f_0 \to \pi^0 \pi^0)$. Combining these two values of Br's we obtain $Br(\phi \to f_0 \gamma) = (3.1 - 3.5) \times 10^{-4}$.
  
\subsection{$\phi \to a_0 \gamma \to \eta \pi^0 \gamma$}
\begin{figure}
\begin{tabular}{ccc}
\epsfig{figure=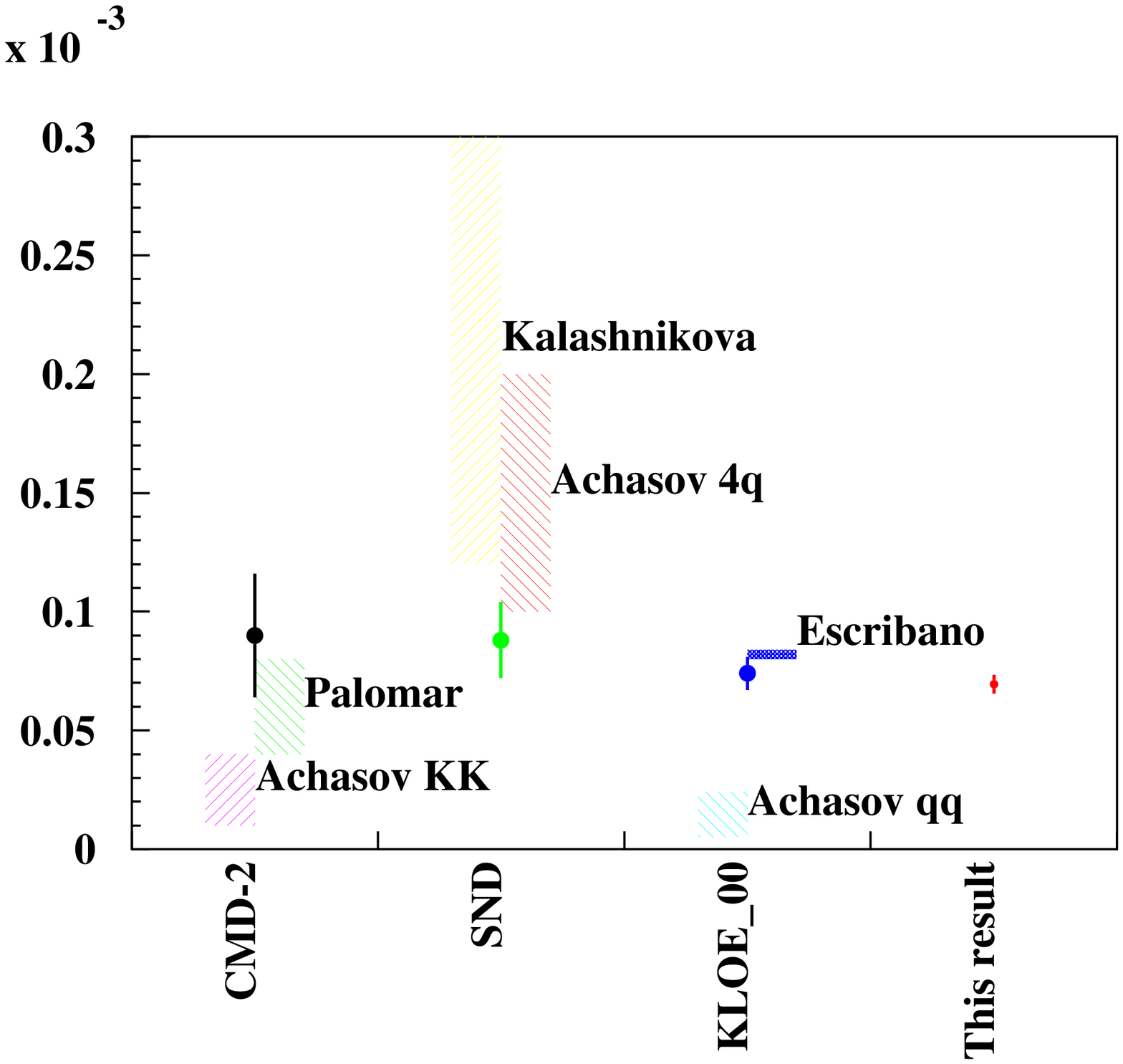,height=2.2 in} & \epsfig{figure=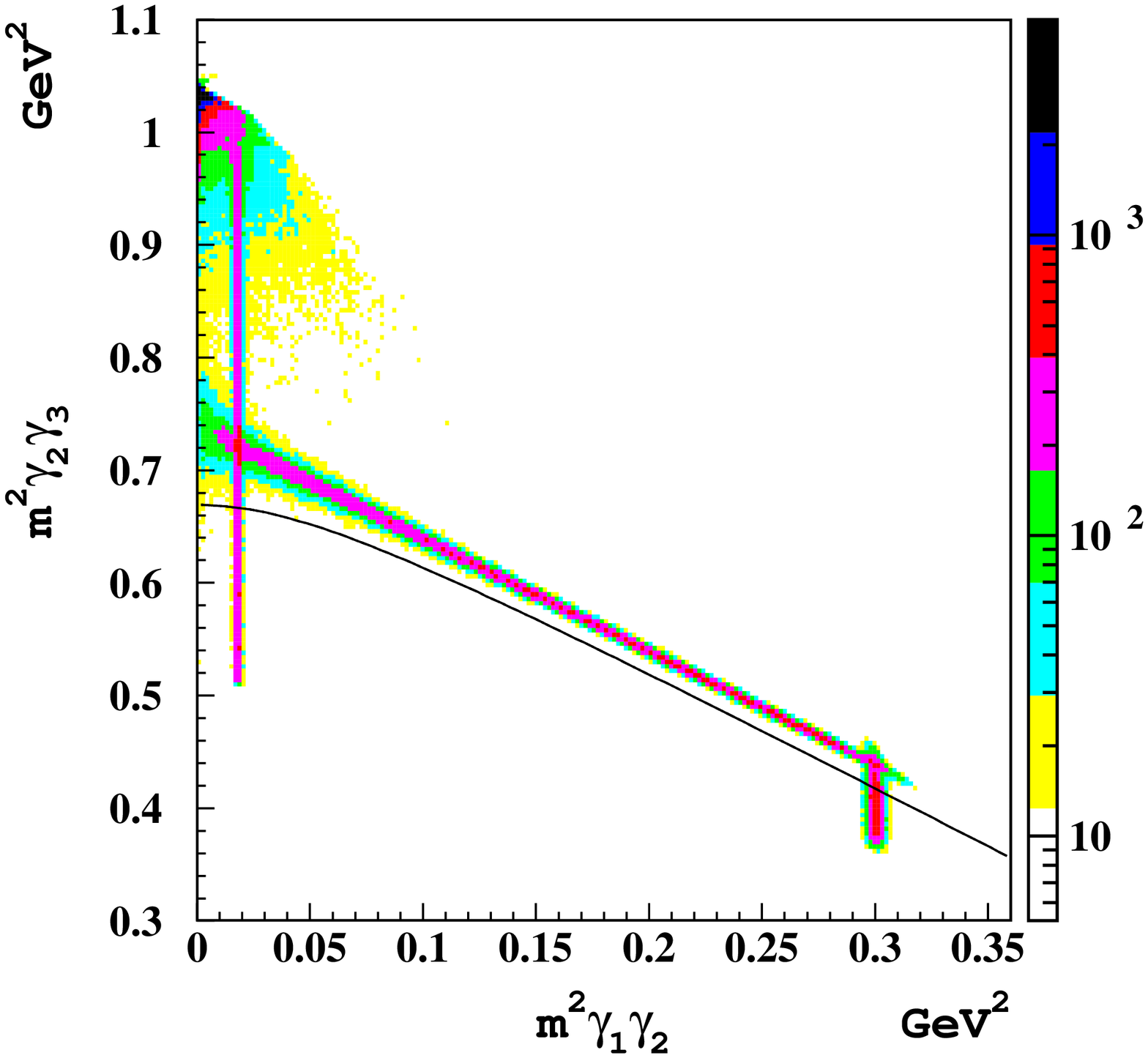,height=2. in} & \epsfig{figure=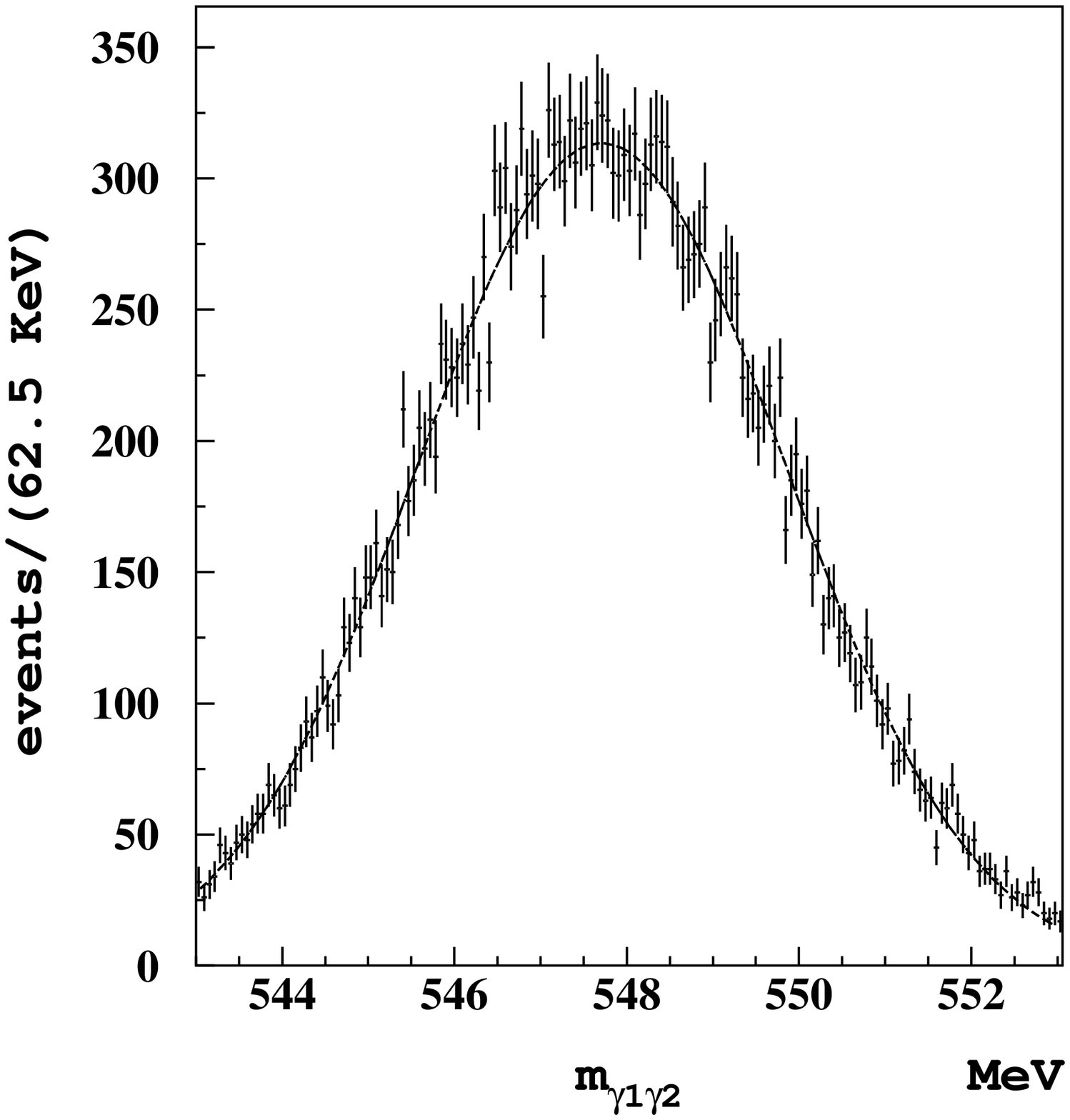,height=2. in} \\
\end{tabular} 
\flushleft
\caption{
\textbf{Left:} Comparison of Br($\phi \to \eta \pi^0 \gamma$) with theoretical models$^{8}$ and previous measurements$^{9}$. 
\hspace{2cm} \textbf{Center:} $3\gamma$ Dalitz plot, \textbf{Right:} $m_\eta$ distribution. \hspace{9cm} }
\label{fig:radish}
\end{figure}

The $\eta \pi^0 \gamma$ final state has only one interfering background coming from the decay chain 
$\phi \to \rho \pi^0 \to \eta \gamma \pi^0$. The contribution of this decay channel is  very small due to the small 
branching ratio $\rho \to \eta \gamma$. In this case a direct
background subtraction is possible in order to extract the $Br(\phi \to \eta \pi^0 \gamma)$. 
Other large not interfering background is present in the final selection mainly coming from 
$e^+ e^- \to \omega \pi^0$ with  $\omega \to \pi^0 \gamma$,  $\phi \to f_0 \gamma$ with $f_0 \to \pi^0 \pi^0$ and $\phi \to \eta \gamma$ with $\eta \to 3 \pi^0$ when 2 photons are lost in the low polar angle region or one or two pairs of photons overlap in a single energy deposit. These background contributions are determined with the use of the MC simulation, after a careful reweighting of each single contribution using background enriched control samples. \\
The number of events that pass the selection after the background subtraction is $N_{sig} = 13099 \pm 172$, the data analysed correspond to an integrated luminosity of 
$L = 413.0 \pm 2.5$ pb$^{-1}$. The efficiency of the whole selection chain is $\epsilon = 37.9 \%$, using $\sigma_\phi = 3090 \pm 80$ nb and $Br(\eta \to \gamma \gamma) = 39.38 \pm 0.26 \%$, we obtain:
\[
Br(\phi \to \eta \pi^0 \gamma) =  \frac{N_{sig}}{\sigma_{\phi} L Br(\eta \to \gamma \gamma) Br(\pi^0 \to \gamma \gamma) \epsilon} =  (6.95 \pm 0.09_{stat.} \pm 0.24_{syst.}) \times 10^{-4}
\]
The systematic error is mainly due to the knowledge of the $\phi$ cross section, the background subtraction and the knowledge of the photon efficiency. 
In fig.\ref{fig:radish}, a comparison of the value of this branching ratio with several theoretical models is shown, together with the previous measurements.

\section{$\eta,\eta'$ mixing angle.}

KLOE has measured \cite{Camilla}  the following ratio $R$ of branching ratios:
\[ 
R \equiv \frac{BR(\phi \to \eta^{\prime} \gamma)}{BR(\phi \to \eta \gamma)} = (4.77 \pm 0.09_{stat.} \pm 0.19_{syst.})\times 10^{-3}
\]
Following the reference by Bramon {\it et al.}\cite{Bra} we can write the $\eta/\eta'$ wave function as a linear combination of non strange $q\bar{q}>$, strange $|s\bar{s}>$ quark pairs plus,
only for  $\eta'$, $|glue>$:
\begin{eqnarray*}
|\eta'>  & = &  cos(\varphi_G)sin(\varphi_p)|q \bar{q}> +  cos(\varphi_G)cos(\varphi_P)|s \bar{s}> + sin(\varphi_G) |glue> \\
|\eta>  & = & cos(\varphi_p)|q \bar{q}>  - sin(\varphi_p)|s \bar{s}>.
\end{eqnarray*}
The ratio $R$ can be related to these parameters using the formula:
\[
R = cot^{2} (\varphi_{P}) cos^2 (\varphi_{G}) \left(
1-\frac{m_s}{{\bar m}}\frac{Z_{NS}}{Z_{S}}\frac{tan \varphi_V} {sin2\varphi_{P}}\right )^2
\left( \frac{p_{\eta^{\prime}}}{p_{\eta}}  \right )^3
\]
we fit this value together with the available data on $\Gamma(\eta' \to \gamma \gamma)/\Gamma(\pi^0 \to \gamma \gamma)$, 
$\Gamma(\eta' \to \rho \gamma)/\Gamma(\omega \to \pi^0 \gamma)$ and $\Gamma(\eta' \to \omega \gamma)/\Gamma(\omega \to \pi^0 \gamma)$ using the 
theoretical estimates for $Z_{NS}$ and $Z_{S}$\cite{ESCRZ}. We obtain $\varphi_P = (39.7 \pm 0.7)^\circ$ and $Z^2_{\eta'} = sin^2{\varphi_G} = 0.14 \pm 0.04$, with $P(\chi^2) = 49$\%. 
Imposing $\varphi_G = 0$ the probability   $P(\chi^2) = 1$ \% and the value $\varphi_P = (39.7 \pm 0.7)^\circ$.

\section{Measurement of the $\eta$ mass.}
The value of the $\eta$ mass has been recently measured with high precision by two collaborations NA48\cite{NA48} ($m_{\eta} = 547.843 \pm 0.030 \pm 0.041$ MeV/c$^2$) 
and GEM\cite{GEM} ($m_{\eta} = 547.311 \pm 0.028 \pm 0.032$ MeV/c$^2$) using different techniques and production reactions. The two measurements differ by more than eight standard 
deviations from each other. The GEM measurement is in agreement with the older ones\cite{PDG} while the NA48 measurement is higher. 
For this reason it is interesting to provide a further measurement of comparable precision in order to clarify the experimental situation.
We measure the mass studying the decay $\phi \to \eta \gamma, \eta \to \gamma \gamma$. A kinematic fit is performed imposing the 4 constraints 
given by the energy-momentum conservation. 
Since the photons are just three the fit overconstrains the energies of the photons that are, practically, determined by the position of the 
clusters in the calorimeter. The inputs of the fit are the energy, the position and the time of the calorimeter clusters,
the mean position of the $e^+ e^-$ interaction point, the total four-momentum of the colliding $e^+ e^-$. Each of these variables is
determined run by run using $e^+ e^- \to e^+ e^-$ events. The absolute $\sqrt{s}$ scale has been calibrated against the $m_{\phi}$ value as measured by CMD-2\cite{CMD2s}. The $\phi \to \eta \gamma$ events are selected by requiring three energy deposits in the calorimeter
and a loose cut on the $\chi^2$ of the kinematic fit. The events surviving the cuts are shown in fig.\ref{fig:radish}, center where the
Dalitz plot for $(m^2_{\gamma _1 \gamma_2}, m^2_{\gamma _2 \gamma_3})$ is shown. Three bands are clearly visible. The band at low $m^2_{\gamma \gamma}$
is given by the $\phi \to \pi^0 \gamma, \pi^0 \to \gamma \gamma$, while the other two bands are $\phi \to \eta \gamma, \eta \to \gamma \gamma$ events.
 With the shown cut in the Dalitz plot we select a pure sample of $\eta, \pi^0 \to \gamma \gamma$ events. The resulting $m_{\gamma \gamma}$ spectrum (fig.
\ref{fig:radish}, right) can be well fitted with a single gaussian with $\sigma \sim 2.1 MeV/c^2$ . 
In order to evaluate the systematic error we have evaluated the uncertainities on all the variables that are given as input to the fit. \\
To determine these uncertainities we have used as a control sample the  $e^+ e^- \to \pi^+ \pi^- \gamma$ events, which allows to check the 
vertex position, the energy response of the calorimeter, 
the alignment of the calorimeter with the Drift Chamber. A small correction to the value of the mass has been found  due to  the kinematic fit algorithm. This correction has been evaluated using the 
MC simulation and half of the correction has been taken as systematic error.
Finally the stability of the result respect to different orientations of the 3$\gamma$ plane has been checked, and the variation taken as systematic error. The various contributions to the systematic error are summarized in table \ref{tab:systematics}.
\begin{table}[t]
\caption{Relative contributions to $m_{\eta}$ systematic error.} \label{tab:systematics}
\vspace{0.4cm}
\begin{center}
\begin{tabular}{|c|l|c|l|}
\hline
Calorimeter response & 2 \% & Vertex position & 1 \% \\
\hline
Angular stability & 26 \% & Fit bias & 70 \% \\
\hline
\end{tabular}
\end{center}
\end{table}
 The result is $m_{\eta} = 547.822 \pm 0.005_{stat.} \pm 0.069_{syst.}$ MeV/c$^2$. In order to check the validity of the whole
precedure, the mass of the $\pi^0$ has been measured 
with the same procedure obtaining $m_{\pi^0} = 134.915 \pm 0.011_{stat.} \pm 0.058_{syst.}$ MeV/c$^2$. This value is in agreement at 1 $\sigma$ level 
with the PDG value $m_{\pi^0} = 134.9766 \pm 0.0006$ MeV/c$^2$. Our preliminary result is in very good agreement with the NA48 measurement ($0.24 \sigma$) and disagrees with GEM by more than 6$\sigma$.

\end{document}